\documentclass[reprint,superscriptaddress,aps,prx]{revtex4-2}

\usepackage[usenames,dvipsnames]{xcolor}
\usepackage[utf8]{inputenc}
\usepackage{graphicx,psfrag}
\usepackage{amssymb}
\usepackage{lineno}
\usepackage[english]{babel}
\usepackage{amsmath}

\usepackage{booktabs} 
\usepackage{multirow} 
\usepackage{floatflt}

\usepackage{soul,xcolor}
\usepackage[colorlinks=true,linkcolor=Blue,citecolor=Blue,urlcolor=Blue]{hyperref}

\usepackage[caption=false]{subfig}
\usepackage[capitalise]{cleveref}

\begin{document}

%\title{Topological analysis of brain signals reveal signatures of epileptogenic dynamics} % in Zebrafish
%\title{Topological analysis of brain signals reveal signatures of epileptogenic dynamics} % in Zebrafish
\title{Topological analysis of brain dynamical signals indicates signatures of seizure susceptibility} % in Zebrafish

%\title{Topological signatures of brain signals reveal precursors of epileptogenic dynamics} % in Zebrafish
%\title{Topological signatures of brain signals reveal precursors of epileptogenic dynamics} % in Zebrafish

\author{Maxime Lucas}
\thanks{\raggedright{Corresponding authors: alexander.skupin@uni.lu; maxime.lucas.work@gmail.com; daniele.proverbio@unitn.it.}}
\affiliation{Department of Mathematics \& Namur Institute for Complex Systems (naXys), Université de Namur, Namur, Belgium}
\affiliation{CENTAI Institute, Turin, Italy}

\author{Damien Francois}
\affiliation{Interdisciplinary Centre for Security, Reliability and Trust; University of Luxembourg; 29, Avenue J.F Kennedy, 1855; Luxembourg}
\affiliation{Luxembourg Centre for Systems Biomedicine; University of Luxembourg; Belvaux, 6 Avenue du Swing, 4367; Luxembourg}

\author{Laurent Mombaerts}
\affiliation{Luxembourg Centre for Systems Biomedicine; University of Luxembourg; Belvaux, 6 Avenue du Swing, 4367; Luxembourg}

\author{Cristina Donato}
\affiliation{Luxembourg Centre for Systems Biomedicine; University of Luxembourg; Belvaux, 6 Avenue du Swing, 4367; Luxembourg}

\author{Alexander Skupin}
\thanks{\raggedright{Corresponding authors: alexander.skupin@uni.lu; maxime.lucas.work@gmail.com; daniele.proverbio@unitn.it.}}
\affiliation{Luxembourg Centre for Systems Biomedicine; University of Luxembourg; Belvaux, 6 Avenue du Swing, 4367; Luxembourg}
\affiliation{Department of Physics and Material Science ; University of Luxembourg; Luxembourg, 162a, avenue de la Faïencerie, 1511; Luxembourg}
\affiliation{National Center for Microscopy and Imaging Research; University of California San Diego; La Jolla, Gilman Drive, CA, 9500; United States}

\author{Daniele Proverbio}
\thanks{\raggedright{Corresponding authors: alexander.skupin@uni.lu; maxime.lucas.work@gmail.com; daniele.proverbio@unitn.it.}}
\affiliation{Department of Industrial Engineering; University of Trento; Trento, 9 via Sommarive, 38123; Italy}

% Potential venues: Nature Neuroscience, PNAS, Nature Communications, Science Advances, NeuroImage, PLoS Computational Biology, Seizure

%\date{December 2023}

\begin{abstract} 
Epilepsy is known to drastically alter brain dynamics during seizures (ictal periods), but its effects on background (non-ictal) brain dynamics remain poorly understood.
To investigate this, we analyzed an in-house dataset of brain activity recordings from epileptic zebrafish, focusing on two controlled genetic conditions across two fishlines. 
After using machine learning to segment and label recordings, we applied time-delay embedding and Persistent Homology---a noise-robust method from Topological Data Analysis (TDA)---to uncover topological patterns in brain activity.
We find that ictal and non-ictal periods can be distinguished based on the topology of their dynamics,  independent of genetic condition or fishline, which validates our approach. 
Remarkably, within a single wild-type fishline, we identified topological differences in non-ictal periods between seizure-prone and seizure-free individuals.
These findings suggest that epilepsy leaves detectable topological signatures in brain dynamics even outside of ictal periods. 
Overall, this study demonstrates the utility of TDA as a quantitative framework to screen for topological markers of epileptic susceptibility, with potential applications across species.
\end{abstract}

\maketitle

\section{Introduction}

Epilepsy, a neurological disorder affecting millions of people worldwide \cite{who_epilepsy}, is characterized by recurring seizures (ictal periods) that disrupt normal brain activity. It is a complex and multifaceted disorder that demands interdisciplinary research to understand its causes and mechanisms, develop effective monitoring and treatment strategies, and ultimately improve patients' outcomes. In particular, for patients with established epilepsy, predicting and detecting seizures from time-series recordings is critical. Advances in wearable devices now enable continuous monitoring of brain activity, and algorithms based on signal processing \cite{andrzejak2023seizure} and machine learning \cite{rasheed2020machine} can now identify seizures. In addition, detailed dynamical models of seizure onset can reproduce common patterns of seizures in various types of epilepsy \cite{Jirsa2014a}, and a recent study has revealed cyclic rhythms in seizures \cite{leguia2021seizure}.

%On longer time scales, a lasting challenge is to detect susceptibility to developing epilepsy in individuals without seizures. 
In contrast to predicting the next seizure in individuals with epilepsy, a lasting challenge on longer time scales is to detect susceptibility to developing epilepsy in individuals who are not yet epileptic---and thus do not experience seizures.
Studies have shown that in about 25\% of cases, epilepsy is acquired through an evident cause such as a stroke, a head injury, or a brain tumor \cite{perucca2020genetics}. The causes of the remaining 75\% of cases are still largely unknown. Genetic studies may help explain some of these cases \cite{perucca2020genetics}, but require extensive screening and do not integrate information about brain dynamics. 
A key question in epileptogenesis is thus whether non-ictal brain dynamics contains signatures of epileptic susceptibility. 
A positive answer could open new avenues for detecting latent epileptic conditions and integrating biological insights with dynamical systems analysis.
In this study, we systematically analyze brain activity from a curated in-house dataset using dynamical systems methods to address this question. 
%for monitoring and detection of latent epileptic conditions, and foster the integration of biological blueprints with the analysis of dynamical systems. Hence, we address it by systematically analyzing brain activity of a curated dataset, under the lenses of dynamical systems methods. 

From a dynamical systems perspective \cite{strogatz2014}, brain activity can be viewed as the output of a high-dimensional system, organized into dynamical attractors corresponding to distinct brain states (e.g., seizures or background activity). Several attractors can coexist and transitions between them are triggered by internal or external stimuli~\cite{breakspear2017dynamic, freyer2009bistability, proverbio2023systematic}. The size of an attractor's basin of attraction reflects its ``attractiveness'' and resilience \cite{menck2013basin,proverbio2024bridging}; the larger it is, the more likely the system will end up in that attractor and the less likely it will leave it. In this framework, seizures can be viewed as dynamics within a dynamical attractor, with the frequency of seizure occurrence being linked to the basin size \cite{chu2017predicting}. 
This perspective provides one way of explaining why diverse biological mechanisms---including genetic, infectious, metabolic, or immune factors \cite{perucca2020genetics}---are associated with epileptogenesis: various conditions can concur to shape similar attractor dynamics, destabilizing the “healthy” attractor and making the seizure attractor more reachable \cite{ da2003dynamical, breakspear2006unifying}.

An important aspect of dynamical attractors is their topology, which captures properties such as overall shape, connectivity, and the presence of structural features such as loops or holes. Topological Data Analysis (TDA), as its name suggests, is a field that provides tools to study the topology (or shape) of datasets and time-series recordings \cite{patania2017topological}. TDA is robust to noise \cite{atienza2016separating, expert2019topological, benjamin2024multiscale}, and captures non-linear features of dynamical attractors \cite{xu2021topological}, providing richer information than other methods based on signal processing and state-space reconstruction \cite{osterhage2007nonlinear, maturana2020critical, wilkat2019no}. Moreover, it is interpretable thanks to its underlying mathematical theory, thus overcoming the shortcomings summary statics, machine- and deep-learning \cite{rasheed2020machine, siddiqui2020review, ding2023compact}. TDA is thus well-suited for high-dimensional data, and enables quantitative comparison of topological features often invisible to traditional methods like recurrence analysis or time-frequency analysis.  Overall, in combination with time-delay embedding \cite{takens1981dynamical}, TDA can aptly reconstruct and analyze the topology of attractors from time series, which has made TDA increasingly popular in neuroscience \cite{petri2014homological, Wang2018a, giusti2016twos, ibanez2019topology, sizemore2019importancea, billings2022topological, ren2023dynamic, santoro2024higher} and other fields such as physiology or finance \cite{karan2021time, Gidea2018, Emrani2014}. 

Although TDA has been applied to epilepsy in a handful of studies, they have focused mainly on seizure detection \cite{fernandez2022topological, piangerelli2018topological, Merelli2015}. These studies, often based on clinical patient datasets, suffer from limited control over biological conditions and may lack control cohorts. To address these limitations, animal models of epilepsy have been developed \cite{Grone2015a, devinsky2018cross}, providing reproducible and controlled conditions for both microbiological studies and time-series analysis \cite{barkmeier2009animal, wei2021detection}. Zebrafish (\textit{Danio rerio}) have emerged as a valuable model organism for epilepsy research and translational research in particular \cite{hunyadi2017automated, fontana2018developing}. Epileptic susceptibility in zebrafish larvae can arise from genetic mutations or be pharmacologically induced using convulsant drugs. In addition, brain activity can be monitored using local field potential (LFP) recordings \cite{cunliffe2015epilepsy}, and precise knowledge of genetic profiles and laboratory conditions supports the establishment of high-quality, curated, and labeled datasets.

In this study, our objective was to quantitatively characterize the topology of brain activity, across biological conditions, to determine whether it can reveal signatures of epileptic susceptibility.
%to quantitatively characterize epileptic susceptibility, identifying its signatures from time-series of brain activity, 
%subject to various biological conditions. 
To this end, we used an in-house Zebrafish dataset of local field potential (LFP) recordings, measured during seizures and background activity from two fishlines, each represented by a wild type and by a controlled genetic mutation. To characterize the topology of the dynamical attractor associated with each recording, we reconstructed the attractor using time-delay embedding, and then computed three topological metrics from TDA---total persistence, persistent entropy, and persistent Betti numbers \cite{smith2021topological}. First, we validated our approach by showing that these topological metrics can discriminate between seizures and background, regardless of fishline and genetic condition. Second, we found that they cannot distinguish between fishlines except for a specific case: background activity of mutants, by looking at the homological dimension 1. This suggests some common topological features of the dynamics between fishlines. Finally, we showed that within a single fishline, individuals that have had seizures, and are prone to develop new ones, can be distinguished from those who have not, solely based on the topology of their dynamics. 
Our findings highlight TDA’s capacity to detect and characterize ictal and non-ictal states, and suggests that it could be used to detect seizure susceptibility by looking for altered dynamical landscapes. 

\section{Results}

\subsection{Experimental Data}

\begin{figure*}
    \centering
    \includegraphics[width=1\linewidth]{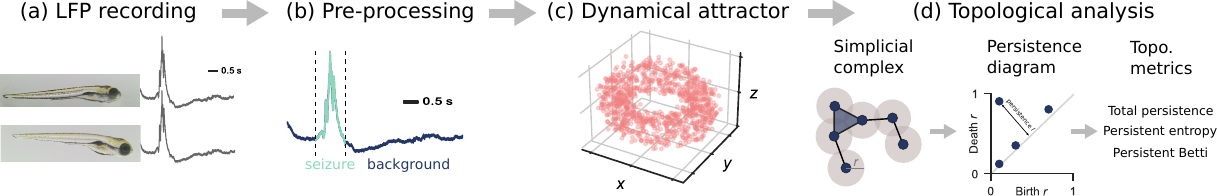}
    \caption{\textbf{Analysis pipeline.} 
    \textbf{(a)} We recorded Local Field Potential activity of Zebrafish larvae from two different fishlines. 
    \textbf{(b)} Each raw LFP recording was pre-processed, fed to a Machine Learning classifier for labeling (\textit{i.e.} distinguishing ``seizure" and ``background", further confirmed by a trained neuroscientist), and split. 
    \textbf{(c)} From each single-label recording, we reconstructed the associated dynamical attractor using time-delay embedding: the recording is mapped from time series in a suitable higher dimensional space. 
    \textbf{(d)} From each dynamical attractor, we compute a persistence diagram using Topological Data Analysis. From it, we extract three topological metrics: total persistence, persistent entropy and persistent Betti numbers.
    Finally, we compare the topology of brain activity by comparing these three metrics under different biological conditions: fishline, genetic mutation, and seizure or background activity.} 
    \label{fig:1}
\end{figure*}

Experiments were carried out on two epileptogenic fishlines that carry mutations (MUT) in either ash1l or kcnq5a genes, as well as their wild-type (WT) variants. All strains can exhibit seizure and non-seizure events. Local field potential (LFP) was recorded with a single electrode per fish (Fig. \ref{fig:1}a). After segmenting the LFP recordings and labeling them, each time series is thus associated with a triplet (Fishline, Condition, Event), where each entry can take two values (\{kcnq5a; ash1l\}, \{MUT; WT\}, \{background; seizure\}). In addition, we benchmarked our results with a fifth strain, a seizure-free wild type (sfWT) ash1l strain. In total, after discarding 14 experiments out of 109 (due to experimental disturbances, see Methods), we ended up with 488 time series from seizures, and 661 from background activity (more details in Table \ref{tab:summaryData}). The number of segmented background events is in general higher than the number of seizures since some anomaly was detected but not confirmed to be an ictal event (see Methods). Note that there are more experiments for the kcnq5a fishline, leading to class imbalance: we took this into account in the statistical analysis. More details on the data collection and pre-processing can be found in Methods.

\begin{table}[htb]
    \addtolength{\tabcolsep}{+0.1cm}
    \centering
    \caption{\textbf{Data summary.} For each epileptogenic fishline, either in variant MUT or WT, we report the total number of experiments performed $N_{\rm exp.}$, the number of experiments discarded from the analysis $N_{\rm drop.}$) (see Methods). For accepted experiments, $N_{\rm acc.} = N_{\rm exp.} - N_{\rm drop.}$, we report the total number of seizures $N_{\rm seiz.}$ and background recordings $N_{\rm backgr.}$, and the mean number of seizures $\left< n_{\rm seiz.} \right> = N_{\rm seiz.} / N_{\rm acc.}$. Seizure-free strain WT does not have seizures (-) and is denoted sfWT.}
    \label{tab:summaryData}
    \begin{tabular}{llrrrrr}
        \toprule
        %Fishline & Condition & No. Exps & No. dropouts & No. seizures & No. backgrounds &  Avg no. seizures / exp. \\
        Fishline & Cond. & $N_{\rm exp.}$ & $N_{\rm drop.}$ & $N_{\rm seiz.}$ & $N_{\rm backgr.}$ & $\left< n_{\rm seiz.} \right>$ \\
        \midrule
        % ash1l & MUT & 20 & 2 & 67 & 83 &  3.7 \\
        % ash1l & WT & 3 & 0 & 22 & 25 &  7.3 \\
        % kcnq5a & MUT & 37 & 11 & 229 & 230 &  8.8 \\
        % kcnq5a & WT & 43 & 1 & 170 & 212 &  4.0 \\
        % kcnq5a & sfWT & 6 & 0 & - & 111 &  - \\
        ash1l & MUT & 20 & 2 & 46 & 54 &  3.7 \\
        ash1l & WT & 3 & 0 & 20 & 24 &  7.3 \\
        kcnq5a & MUT & 37 & 11 & 223 & 224 &  8.8 \\
        kcnq5a & WT & 43 & 1 & 56 & 63 &  4.0 \\
        kcnq5a & sfWT & 6 & 0 & - & 111 &  - \\
        \midrule
        \multicolumn{2}{l}{Total} & 109 & 14 & 488 & 661 &  6.0 \\
        \bottomrule
    \end{tabular}
\end{table}

%

%During the experiments, each LFP time series was recorded with a single electrode per fish. The experiments were first carried out on two epileptogenic fishlines carrying mutations in ash1l or kcnq5a genes and corresponding  wild type variant which can exhibit background and seizure events. Post-processed time series windows, which are subsequently analysed using TDA, is thus associated with three labels (fishline, variant, event) with two possible values each. Each fishline is thus labelled as $x$-MUT and $x$-WT, where $x$ is substituted by the corresponding strain (see Table \ref{tab:summaryData}). For each of them, we differenntiate between background and seizure events. In addition, we benchmarked our results with a fifth strain, a seizure-free wildtype (WT). These time series were pre-processed as detailed in Methods, until obtaining a collection of curated windows. Note that, out of all performed experiments, some were dropped out, when the experimenter had noted that some experimental issue had occurred (like ``electrodes not well-placed''). 

%The full list of experiments for each fishline, their corresponding dropouts, and the statistics about seizure and background events, are listed in Table \ref{tab:summaryData}, which summarises the data set. After all pre-processing steps, there was a total of 488 seizures and 661 background events. 

\subsection{Analysis pipeline}

\Cref{fig:1} illustrates the complete analysis pipeline, which consists of four main steps: (a) experimental recording of brain activity through LFP under different biological conditions, (b) pre-processing and automated labeling of the recordings into ictal and non-ictal periods, before splitting them into single-label recordings, (c) reconstruction of the dynamical attractor associated to each labeled recording using time-delay embedding, and (d) quantification of attractor topology using persistent homology. We then compared the topology of the attractors under different conditions. Further details can be found in the Methods section.

After experimental recording and segmentation of background and seizure activities (\cref{fig:1}a,b), we reconstructed the higher-dimensional dynamics of fish brains to analyze the topology of the recorded LFP dynamics. Using \textit{time-delay embedding}, each single-electrode time series  $\{x(t_0), x(t_1), \dots, x(t_N) \}$ was transformed into a $d$-dimensional time series by mapping each point $x(t_i)$ to a vector $\mathbf{x}(t_i) = [x(t_i), x(t_i+\tau), \dots, x(t_i+(d-1)\tau)]$, where $\tau$ is the delay and $d$ is the embedding dimension (\cref{fig:1}c). According to Takens' theorem, this embedding preserves the topology of the underlying attractor, allowing us to represent the LFP dynamics as a set of points in $d$-dimensional space, known in dynamical systems as a dynamical attractor (see Methods).

We analyzed the topology of each attractor using \textit{persistent homology}, a method that constructs simplicial complexes (generalization of networks) from a set of data points and analyzes its topological features across multiple scales. These topological features can represent connected components of the simplicial complex (homological dimension 0, $H_0$) or holes (homological dimension 1, $H_1$) or higher-dimensional ``holes'' in higher homological dimensions. These features are captured in a \textit{persistence diagram} (\cref{fig:1}d), which encodes the birth and death of these topological features $(r_B, r_D)$, where $r_B$ is the birth and $r_D$ the death scale of a feature. The persistence of a feature, say a hole, is given by $r_D - r_B$---a large persistence indicates that this hole is persistent across many scales and is hence a prominent topological feature. Visually, the persistence of a topological feature is its distance to the diagonal (\cref{fig:1}d). Conversely, points close to the diagonal have low persistence and are often considered noise.

We then computed three topological metrics to summarize each diagram: total persistence, persistent entropy, and persistent Betti numbers (see Methods for formal definitions).
Total persistence is the sum of persistences of all topological features in the persistence diagram. A large total persistence indicates marked topological features that are persistent across scales. Persistent entropy is a measure of the diversity in the feature's persistences, and large persistent entropy indicates diverse persistences. The persistent Betti number in homological dimension $k$ counts the number associated topological features across scales, that is, the number of points in the persistence diagram. A large persistent Betti number indicates a large number of topological features.
We compared these three metrics to assess differences between fishlines, genetic variants, and event types.

\subsection{Topology of LFP dynamics}

To begin with, we assessed that TDA was effective to detect and differentiate topological signatures. In fact, the data-driven reconstruction of LFP state-spaces requires further quantification and systematic categorization to be automatically processed and segmented. TDA is, in principle, an effective method to perform such task, but its statistical power in discriminating seizure events and other topological blueprints still needs assessment.

We thus tested whether there were statistically significant differences in topology by performing a permutation ANOVA with three factors: Fishline (ash1l vs kcnq5a), Condition (mutant vs wild type), and Event (background vs seizure). We did this for each topological metric and each homology dimension separately. For this analysis, we used the complete dataset summarized in \cref{tab:summaryData} except for the seizure-free group (ash1l sfWT). The results are described below and summarized in \cref{tab:anova}. Overall, Event was a significant main effect with a large effect size in all cases, whereas Condition and Fishline are main effects (or their interaction is significant) in much fewer cases and with much smaller effect sizes. 

First, we examined $H_0$, which represents connected components in the attractor. At $H_0$, for total persistence, permutation ANOVA indicated that there was no main effect of any of the factors or their pairwise interactions. Only their triple interaction was significant but with a very small effect size ($p = 0.02$, $\eta^2_p = 0.004$).
For both persistent entropy and persistent Betti numbers, the Event factor was the only main effect with a large effect size ($p < 10^{-15}$, $\eta^2_p = 0.58$). The triple interaction was also significant for both metrics ($p = 0.010$ and $p < 10^{-15}$, respectively), but with very small effect sizes ($\eta^2_p = 0.007$ and $\eta^2_p = 0.008$, respectively).
This indicates that, at the level of connected components ($H_0$), the topology differed significantly between background and seizures in two of the three metrics, explaining approximately 60\% of the variance in the data. However, the topology did not differ between fishlines or genetic conditions.

\begin{table}[t]
    \setlength{\tabcolsep}{4pt}
    \centering
    \caption{\textbf{Statistical analysis summary.} Significant effects on the topological metrics Total Persistence, Persistent Entropy, and Persistent Betti, determined by a three-way permutation ANOVA with factors Event (background vs seizure), Condition (MUT vs WT), and Fishline (kcnq5a vs ash1l), for homological dimensions $H_0$ and $H_1$. We report the $p$-value associated to each effect, and the effect size measured by partial eta squared $\eta^2_p$.}
    \begin{tabular}{p{0.5cm}p{2cm}lrr}
    
    \cmidrule[\heavyrulewidth]{2-5}
     & \textbf{Metric} & \textbf{Signif. effects} &  \textbf{$p$-value} & \textbf{$\eta^2_p$ (\%)} \\
    \midrule
    \multicolumn{4}{l}{$H_0$}\\
    \midrule
    & \multicolumn{2}{l}{Total Persistence} & & \\
    & & Triple Interaction & 0.02 & 0.4 \\
    \cmidrule(r){3-5}
    & \multicolumn{2}{l}{Persistent Entropy} & & \\
    & & Event & $<10^{-15}$ & 58.0 \\
    & & Triple Interaction & 0.010 & 0.7 \\
    \cmidrule(r){3-5}
    & \multicolumn{2}{l}{Persistent Betti} & & \\
    & & Event & $<10^{-15}$ & 58.0 \\
    & & Triple Interaction & $<10^{-15}$ & 0.8 \\
    \midrule
    \multicolumn{4}{l}{$H_1$ } \\
    \midrule
    & \multicolumn{2}{l}{Total Persistence} & & \\
    & & Event & $<10^{-15}$ & 24.2 \\
    & & Condition & $<10^{-15}$ & 1.4 \\
    & & Condition:Fishline & 0.031 & 0.5 \\
    & & Triple Interaction & $<10^{-15}$ & 1.3 \\
    \cmidrule(r){3-5}
    & \multicolumn{2}{l}{Persistent Entropy} & & \\
    & & Event & $<10^{-15}$ & 66.6 \\
    & & Fishline & 0.045 & 1.3 \\
    & & Condition:Fishline & 0.030 & 0.6 \\
    \cmidrule(r){3-5}
    & \multicolumn{2}{l}{Persistent Betti} & & \\
    & & Event & $<10^{-15}$ & 74.4 \\
    & & Fishline & $<10^{-15}$ & 2.0 \\
    & & Condition:Fishline & 0.033 & 0.7 \\
    & & Triple Interaction & $<10^{-15}$ & 1.2 \\
    \bottomrule
\end{tabular}
\label{tab:anova}
\end{table}

Second, we examined $H_1$, which represents cycles or ``holes" in the dynamical attractor (like in Fig. \ref{fig:1}e). For total persistence, permutation ANOVA indicated a significant main effect for both Event, with a large effect size ($p < 10^{-15}$, $\eta^2_p = 0.242$), and Condition, but with a small effect size ($p < 10^{-15}$, $\eta^2_p = 0.014$). There was also a significant Condition:Fishline
interaction ($p = 0.031$, $\eta^2_p = 0.005$) and triple interaction ($p < 10^{-15}$, $\eta^2_p = 0.013$), both with small effect sizes.
For persistent entropy, Event was a significant main effect with a very large effect size ($p < 10^{-15}$, $\eta^2_p = 0.666$). Fishline was also a significant main effect, but with a small effect size ($p = 0.045$, $\eta^2_p = 0.013$). The Condition:Fishline
interaction was significant, but with a very small effect size ($p = 0.030$, $\eta^2_p = 0.006$).
The results for persistent Betti numbers were similar to those for persistent entropy, as observed for $H_0$. Both Event and Fishline were significant main effects ($p < 10^{-15}$ in each case) with a very large effect size for Event ($\eta^2_p = 0.744$) but a small effect size for Fishline ($\eta^2_p = 0.02$). There was again a significant Condition:Fishline
interaction ($p = 0.033$, $\eta^2_p = 0.007$) and a significant triple interaction ($p < 10^{-15}$, $\eta^2_p = 0.012$), both with small effect sizes.
Similarly to $H_0$, these results indicate that, at the level of cycles, the topology differs significantly between the background and seizures, explaining between 20\% and 75\% of the variance in the data, depending on the topological metric. In particular, persistent entropy and the persistent Betti number perform better by explaining substantially more variance from the Event factor, both in $H_0$ ($\eta^2_p=58\%$ vs no main effect) and $H_1$ ($\eta^2_p=66$-$75\%$ vs $\eta^2_p=24.2\%$).

Overall, these results indicate that the topology of the LFP dynamics is affected by the three biological factors under study (the fishline, the genetic condition, and the type of event) in a complex and intertwined pattern. The topological metrics that we used allow us to uncover this pattern. 
%Importantly, the topological indicators that we use reveal t  these results confirm that different TDA metrics are consistent in discriminating topological features, and that the topology of LFPs primarily differs between seizure and background activities, whereas there are shared characteristics when comparing fishlines and conditions that are seizure-prone. 
%This analysis also allows us to extract meaningful interpretations by accounting for the different inner variability within fishlines and the imbalance among classes (see Supplementary Material for further details). 
In the following, we focus on three aspects of these results by performing post hoc analyses.

\subsection{Seizures and background dynamics are topologically different}

\begin{figure}[htb]
    \centering
    \includegraphics[width=.99\linewidth]{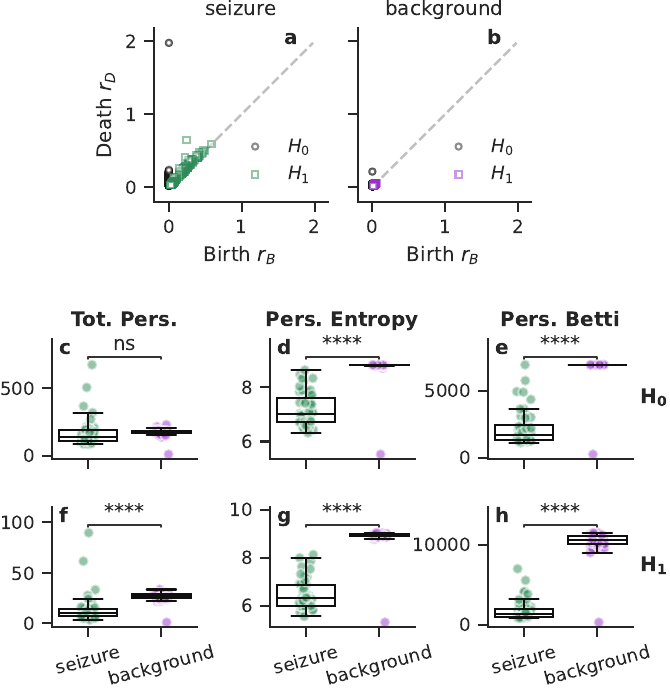}

    \caption{\textbf{The topology of dynamics discriminates between seizures and background LFP activity.} Example of persistence diagrams for a mutant of the ash1l fishline during (a) seizure and (b) background activity. We show three topological metrics---(c, f) total persistence, (d, g) persistent entropy, and (e, h) persistent Betti---for seizures and background, in homological dimensions (c-e) 0 and (f-h) 1. Pairwise Welch's $t$-test indicate a significant difference in all cases with $p<10^{-15}$ (``****''), except for total persistence in $H_0$ where the difference is not significant (``ns''). Here, results are shown for mutants of the ash1l fishline---other fishlines and mutations are reported in \cref{fig:s1}.}
    \label{fig:2}
\end{figure}

%After reconstructing the dynamical attractor of LFP activity,
The strongest signal---by far---in the above analysis is that we can automatically discriminate between seizures and background activity based on the topology of the dynamics, as indicated by the factor Event being a main effect with high variance explained by almost all topological metrics. In \cref{fig:2}, we illustrate this for the mutants of the ash1l fishline : the distribution of topological metrics differ significantly between seizures and background (as shown by pairwise Welch's $t$-tests, $p<10^{-15}$)---except for total persistence in $H_0$---and the three topological metrics are consistently higher in background than in seizures. Pairwise Welch's $t$-tests yield similar results for the other fishline and genetic mutation, as reported in the Supplementary Material \cref{fig:s1}.

Overall, these results confirm that the dynamical attractors associated with ``healthy'' and ``epileptic'' activity are altered, irrespective of the underlying biological condition, and that TDA is an appropriate tool to discriminate between them.

% \bigskip

% \hrule

% Results description accompanying each of the key figures

\subsection{Different fishlines share common topological features of LFP dynamics under most conditions}
%\subsection{Different seizure-prone mutants have common topological features of brain dynamics}

A natural next question is whether the topology of the dynamics can allow us to discriminate between fishlines. The answer to this question is not as clearcut as it was in the case of the even type in the previous section. We found that under most biological conditions and topological metrics, the fishlines can not be distinguished. However, going to homological dimension 1 ($H_1$) and looking at mutants, the fishlines do differ significantly, although with small effect sizes. We detail these results below. 

\begin{figure}[t]
    \centering
    \includegraphics[width=.99\linewidth]{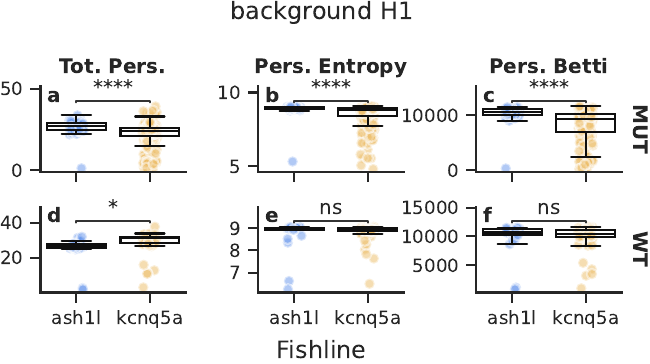}
    \caption{\textbf{The topology of dynamics can consistently discriminate between fishlines only under a very specific lens: background activity of mutants by looking $H_1$.} We show three topological metrics---(a, d) total persistence, (b, e) persistent entropy, and (c, f) persistent Betti---for two genetic conditions---MUT and WT. The homological dimension is 1 and we show background activity (see \cref{fig:s2} for seizures). Pairwise Welch's $t$-test indicate a significant difference in mutants for all three metrics with $p<10^{-15}$ (``****''). ``ns'' denotes no significant differences and ``*'' indicates $10^{-2} <p < 5  \times 10^{-2}$.}
    \label{fig:3}
\end{figure}

At the level of connected components, $H_0$, none of the topological metrics can discriminate between fishlines---indeed, the permutation ANOVA told us that Fishline was not a main effect. However, at the level of cycles, $H_1$, Fishline is a main effect for persistent entropy and the persistent Betti number, and there is an interaction Condition:Fishline, as reported in the last section. To better understand this effect, we performed post hoc analyses in the form of pairwise Welch's $t$-test between the two fishlines, for each (Condition, Event) subgroup. In \cref{fig:s2}, we report that, for seizures, the two fishlines differ only significantly 
%($p=...$) 
in persistent entropy in MUT, but not in the other two topological metrics, and not in WT. This is a reflection of the Condition:Fishline interaction found by the permutation ANOVA. 
For background activity, we have a more consistent signal: the two fishlines differ significantly in all three metrics in MUT, but only for total persistence in WT (\cref{fig:3}).
In summary, we were able to discriminate between fishlines only under a very specific lens: only from the background activity of the mutants by looking at cycles ($H_1$)---but not from seizures, not from WT, and not from $H_0$.

Overall, being prone, or having developed seizures, make the two genetically different fishlines very close in the topological space, unraveling a shared dynamics at fundamental level.

\subsection{Detecting seizure-prone specimen from background}

Finally, we asked: Is there any topological difference between the background activity of fish that have had seizures and those that have not?
Unraveling such differences could help identify early signs of seizure susceptibility based only on LFP signals.
To test this, we used the background activity recordings of wild-type kcnq5a fishline, and compared two groups: those with seizures (WT) and the seizure-free ones (sfWT) (see \cref{tab:summaryData}).

\begin{figure}[t]
    \centering
    \includegraphics[width=.99\linewidth]{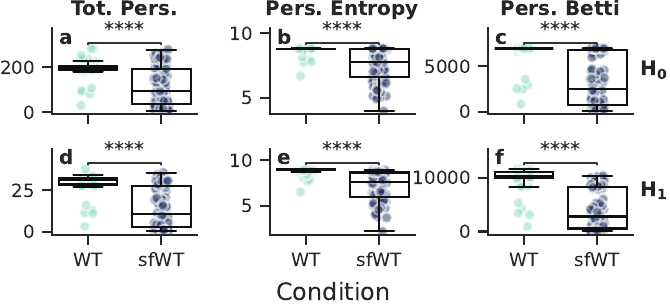}

    \caption{\textbf{Wild types with (WT) and without (sfWT) seizures can be discriminated topologically from background LFP dynamics.} We show three topological metrics---total persistence, persistent entropy, and persistent Betti---for two populations of a single fishline (kcnq5a): those with and without seizures, in homological dimensions 0 and~1.}
    \label{fig:4}
\end{figure}

\Cref{fig:4} shows the comparison between these two groups for the three topological metrics for $H_0$ and $H_1$. All three metrics differ significantly between the WT and sfWT groups, as indicated by pairwise Welch's $t$-tests (all $p<10^{-7}$) showing large effect sizes ($\eta^2$ values between 25-35\% for $H_0$ and between 37-46\% for $H_1$). In all cases, the topological metrics take higher values in WT than in sfWT.  The variance of the sfWT group was also significantly larger in all cases as indicated by a pairwise Levene test (\cref{fig:s3}).
In addition, we observed that the dynamics of the WT and MUT groups were topologically more similar to each other than to the sfWT group (\cref{fig:s4}). Surprisingly, this indicates  that in this case, having had a seizure affects the topology of the brain activity more than having any mutation. Disruptions in the attractor landscape can thus help differentiating individuals that have had seizures (and could have others) from the non-epileptic individuals, and that topological markers may complement or even surpass certain biomarkers.

\section{Discussion}

Our study demonstrates the potential of Topological Data Analysis (TDA) to reveal and characterize the dynamical landscapes of epileptic brain activity.
We systematically analyzed the topology of both ictal (seizure) and non-ictal (background) activity in an in-house dataset of LFP recordings from Zebrafish models for epilepsy. 
These animal models allowed us to investigate the effect of different controlled genetic conditions on the brain activity. By combining time-delay embedding and TDA, we showed topological differences between seizure and background activity, uncovered common features and subtle differences between genetic conditions, and identified seizure susceptibility from normal background brain activity. These findings contribute to a growing body of evidence supporting the hypothesis that seizures emerge from altered dynamical landscapes \cite{breakspear2017dynamic, freyer2009bistability} shaped by underlying biological conditions \cite{da2003dynamical}.

%It has been hypothesized that seizures emerge from dynamical landscapes \cite{breakspear2017dynamic, freyer2009bistability} that are altered by neurological and biological conditions \cite{da2003dynamical}. Our study supports this hypothesis by providing a comprehensive and quantitative analysis of seizure dynamics from a controlled experimental dataset of Zebrafish animal models, which allows to precisely characterize the genetic underpinnings epilepsy.  
%The use of Topological Data Analysis, a stable and powerful method for multiscale analysis of brain dynamics, sheds new insights about the landscape topology emerging from LFP signals. Our results support the use of TDA as an interpretable complement to machine- and deep-learning tools \cite{siddiqui2020review} for data-driven detection of epileptic seizures, and unravel that seizures evolve over a dynamical attractor that display altered patterns from those of background activity. On top of the significance for monitoring applications, this observation supports and informs the use of theoretical models to reproduce attractors and bifurcations in brain activity \cite{breakspear2017dynamic, jirsa2017virtual}.

In our analysis, we were able to discriminate between seizure and background activity based on the topology of the associated dynamical attractors demonstrating the applicability of our approach. Background activity exhibited significantly larger persistent entropy, persistent Betti numbers, and total persistence (except for connected components, $H_0$), regardless of genetic conditions. This indicates that, even though the total ``topological content" is of similar magnitude at the level of connected components, topological features are more abundant and diverse in background activity at the level of connected components and holes. This suggests that background dynamics is characterized by more complex topological structures, while seizure dynamics reflect a reduction in this complexity, possibly due to pathological synchrony or a collapse into lower-dimensional attractors.

%Our study allows to classify between background and seizure states across fish lines and genetic conditions, and to unravel precursor signatures of seizure susceptibility in Zebrafish larvae. We verify that even different genomic profiles yield similar dynamical blueprints for brain dynamics, which, after processing with TDA, may reveal anticipated susceptibility for seizure states. Our results allow to build a robust and quantitative method to identify similar dynamical states, beyond simple metrics like inter-spike interval, enabling comparison across fish lines and, potentially, across species;moreover, it paves the way to develop screening tools for advanced seizure sensitivity detection. %% Write that we also support the idea of using universal dynamical models to characterise epilepsy in terms of landscape disruption, bridging the scales between biological or neurological microstates and emerging dynamics.

We also found that dynamical topology did not discriminate between different fishlines or genetic variants in most cases---not from seizures, wild types, or connected components ($H_0$). This suggests that, for the most part, the brain activity of the epileptic fish shared common topological signatures across fishlines and genetic conditions. The only combination of factors for which we could discriminate between fishlines was in background activity of mutant fishlines, by looking at holes ($H_1$). Here, one fishline (ash1l) exhibited larger total persistence, persistent entropy, and persistent Betti numbers compared to kcnnq5a. In that case, although the differences between fishlines were significant, the effect sizes were small. This suggests that, while epilepsy may induce common topological signatures across species and genetic mutations, subtle differences may still emerge under specific conditions, which can be revealed by TDA. These differences could reflect variations in the way genetic mutations influence brain activity.

The most striking result is the ability of TDA to distinguish between seizure-prone and seizure-free individuals within the same fishline and associated genetic mutation, based solely on background activity. This finding raises two interesting biological hypotheses: either the topological signature found in the background activity of seizure-prone individuals indicates  the following  seizure, or it reflects lasting changes induced by prior seizures. Both scenarios are compelling: the former would suggest that TDA could be used as a predictive tool to detect latent epilepsy before seizures manifest, while the latter points to the plasticity of brain dynamics and the potential for seizures to cause persistent alterations to background brain activity. Further experiments, particularly longitudinal studies, are needed to disentangle these possibilities and explore their implications for early detection and intervention.

A key limitation of our approach towards on-line usage is its computational cost, since measuring persistent homology of dimension $k$ ($H_k$) requires computing simplices up to $k+1$ dimensions. The number of those simplices, for a dataset of $N$ points, grows exponentially $\sim \mathcal{O}(N^{k+1})$. For that reason, most studies using TDA only compute low homological dimensions, e.g. 1 or 2 as in the present study. Homological dimensions 0 and 1, however, are usually fast enough to be promising for real-time applications \cite{fernandez2022topological}.

%Transitioning to applications on human recordings would require assessing the robustness of persistent topological metrics on LFP and EEG signals, potentially recorded through multiple electrodes \cite{billings2022topological}. Extraction of TDA signals from LFP has already been performed \cite{Merelli2015, piangerelli2018topological}, and future studies may further confirm their use for landscape characterization and seizure precursor detection.

%Our analysis proved to be robust for all fish conditions. The few discarded data series were associated to experimental disturbances, like misplaced needles or off-temperature conditions in the kcnq5a fish line, or to uncertain seizure labeling by both the algorithm and the expert. Transitioning to applications on human recordings would require assessing the robustness of persistent topological metrics on LFP and EEG signals, potentially recorded through multiple electrodes. Extraction of TDA signals from LFP has already been performed \cite{Merelli2015, piangerelli2018topological}, and future studies may further confirm their use for landscape characterization and seizure precursor detection.
%Moreover, we extracted TDA metrics from landscapes reconstructed via dynamical embedding from open libraries, which may be sensitive to noise in data. TDA is a robust technique against noise artifacts \cite{atienza2016separating}, but new embedding techniques involving noise filtering may further help reducing the observed variability. 

Even today, a significant percentage of epileptic patients 
%remain undia long time or 
do not fully respond to drug treatment \cite{loscher2020drug}. 
%These challenges suggest that 
Our understanding of underlying epileptogenetic mechanisms is still limited and can result in poor screening capabilities and a lack of specific diagnoses. Quantitative analysis of seizure susceptibility with TDA has the potential to complement the clinical approach to epilepsy.
%Our findings underscore the generality of TDA as a framework for analyzing complex brain dynamics. The noise-robustness and scale-invariance of persistence diagrams make this approach well-suited for studying epilepsy across species and conditions. 
%An accurate signal could provide additional screening capabilities to identify susceptible patients early on and promptly initiate treatment. In addition, it may help forecasting states close to seizure onsets, triggering warnings and interventions. 
It could enhance early screening, enabling the identification of at-risk patients and timely initiation of treatment. 
TDA could also aid in forecasting seizure onsets, complementing existing approaches \cite{krook2013demand, baud2018multi, brinkmann2021seizure}.
%Integrating multi-faceted approaches  for quantifying seizure risks, building on automated TDA-based assessment of seizure precursors and potentially embedded in portable devices \cite{brinkmann2021seizure}, has the potential to improve the long-term monitoring and treatment of people living with epilepsy by complementing other existing approaches.
Translating these methods to human datasets, such as EEG or MEG recordings, would require assessing the robustness of persistent topological metrics on LFP and EEG signal. It could open new avenues for detecting latent epilepsy, monitoring disease progression, or tailoring treatments. Overall, our study demonstrates that, by bridging dynamical systems theory and neuroscience, TDA provides a novel perspective on brain activity that complements traditional analytical approaches.

%In this work, we have processed time series that were already segmented using a machine learning toolbox in combination to expert's judgement. Early warnings for epileptic events, like in finance \cite{akingbade2024topological} 

%TDA much better than Inter Spike Interval (biased and summary statistics) and frequency analysis: more stable, robust to noise and gives global information about attractor, that gives properties to inform the development of models.

\section{Methods}

The analysis pipelines involves several steps (\cref{fig:1}): data collection (from published datasets or from new experiments), data processing for topological data analysis (which includes preprocessing of raw signals, including anomaly detection, automated time series segmentation into background and seizure events, time-delay embedding to reconstruct topological information using dynamical attractors, and TDA), and statistical analysis on TDA results. Each step is detailed in the subsections below.

\subsection{Data collection}

The time series data was generated by measuring Local Field Potential (LFP) recordings in zebrafish (\textit{Danio rerio}) larvae from a KCNQ5 LoF model kcnq5a$^{\text sa9563}$ zebrafish and the ASH1L model ash1l$^{\text sa19097}$ obtained from ZFIN repository. 

\subsubsection{Ethics statement} 

Zebrafish were handled as described previously in \citep{moein_casian_2018, martins_seizure-induced_2023} at the Luxembourg Centre for Systems Biomedicine (LCSB). The Aquatic Facility at the LCSB is registered as an authorized breeder, supplier, and user of zebrafish by the relevant agency of the Government of Luxembourg (Ministry of Agriculture, Viticulture and Rural Development). Experiments using zebrafish larvae at 5 days post-fertilization were performed under Gran-Ducal decrees. All practices involving zebrafish were performed in accordance with European laws, guidelines and policies for animal experimentation, housing, and care (European Directive 2010/63/EU on the protection of animals used for scientific purposes) and following the principles of the Three Rs. Furthermore, we carefully comply with ARRIVE guidelines.

\subsubsection{Zebrafish husbandry}
Adult zebrafish  were maintained in the Aquatic Facility of the Luxembourg Centre for Systems Biomedicine and housed at 28.5 $^\circ$C in a 14-h/10-h light/dark cycle according to standard protocols. Embryos were obtained by natural spawning and fertilized eggs were selected and raised at 28 $^\circ$C in 0.3X Danieau’s medium (17 mM NaCl, 2 mM KCl, 0.12 mM MgSO4, 1.8 mM Ca(NO3)2, 1.5 mM HEPES pH 7.5 and 1.2 $\mu$M methylene blue). Developmental staging was accessed by following standard procedure \cite{kimmel1995stages} and larvae were used used 5 days post fertilization (dpf) for experiments. The corresponding genotype (WT vs heterozygous and homozygous) was accessed by a posteriori sequencing of the corresponding genes kcnq5a and ash1l, respectively. For the analysis, we use homozygous mutants (MUT).

\subsubsection{Local field potential recordings}
LFP were performed as previously described in \cite{martins_seizure-induced_2023}. In brief, each 5 dpf larva was placed in a 50 $\mu$L of Danieau’s medium in the recording chamber with a transfer pipette, and then 200 $\mu$L of 2\% low melting point agar were added. The chamber was transferred on the stage of a stereomicroscope for LFP recordings of a full electrophysiology system (Scientifica SliceScope Pro 1000) equipped with a MultiClamp 700B amplifier and Digidata 1550 A digitizer (Axon instruments, USA). The local field potential (LFP) was recorded at 100 kHz in current clamp mode by a glass microelectrode (4–10 M$\Omega$ resistance) back loaded with extracellular recording solution and placed under visual guidance in the medial tectal band of the midbrain.

\subsubsection{Dataset}
The resulting dataset is a collection of time series for brain activity, along with information about the fishline and information on the mutation type (WT, heterozygous and homozygous). The full data set was first parsed to look for warnings about experimental issues or potential bias, such as electrodes not being well-placed or signals looking altered. This initial pre-processing resulted in a few dropouts for each fishline (Table \ref{tab:summaryData}).

\subsection{Data segmentation and labeling}
The original local field potential (LFP) signal was downsampled from 100kHz to 2000Hz, discarding frequencies higher than 2000Hz which are usually associated with random noise instead of biologically relevant brain dynamics. Then, the 50Hz signal artifact was removed with a notch filter. 

The downsampled recording was then analysed with an automated anomaly detection pipeline. For this, the signal was first decomposed into 11 sub-frequency ranges, through a multi-resolution analysis using the Maximal Overlap Discrete Wavelet Transform (MODWT) and a Daubechie 4 (db4) wavelet.
Wavelet Transforms overcome the main limitation of the Fourier Transform. The latter characterizes the original signal in the frequency domain but loses information on the all-time domain; instead, Wavelet Transform creates a representation of the signal in the time and frequency domain, which allows the localization of time-dependent information in the signal. 
The MODWT is a non-decimating wavelet transform that does not downsample the signal at each scale during processing and produces time-aligned signals. This allows for a straightforward correlation between events in the original signal and in each extracted signal for all sub-frequency ranges. Daubechie 4 wavelets have two vanishing moments, easily encode second-order polynomials, and are widely used to cope with signal discontinuities \cite{vonesch2007generalized}. The 11 sub-frequency ranges were [1000-2000]Hz, [500-1000]Hz, [250-500]Hz, [125-250]Hz, [62-125]Hz, [31-62]Hz (gamma waves), [16-31]Hz (beta waves), [8-16]Hz (alpha waves), [4-8]Hz (theta waves), [2-4]Hz (high delta waves) and [0-2]Hz (low delta waves).

These decomposed signals were then fed to the anomaly detection algorithm, which applies an amplitude threshold in the [62-125]Hz, [31-62]Hz, and [16-31]Hz sub-frequency ranges, as well as a temporal threshold, to select and extract all seizure candidates.
To detect anomalous signals, the algorithm applies a threshold to select only the data points above the 95th percentile as well as a contiguity threshold which filters all anomalies occurring for less than a second.
The amplitude threshold is applied to each targeted sub-frequency range, the temporal threshold is applied to all data points selected through the amplitude threshold in the three sub-frequency ranges simultaneously.

This processed helped to automatize the process and to provisionally assign a ``seizure" labeling to anomalies. Background segments resulted from the time series cropped between anomalies. After extraction, all candidates were submitted for expert evaluation to produce labels identifying anomalies or seizures. Some anomalies did not pass the expert judgment and were discarded from the seizure set. In these cases, segmented background time series were not artificially collated but were kept separate to prevent artifacts. 

%Part of the labeled dataset was used to train a random forest gradient boost algorithm through the production of feature-engineered vectors containing descriptions of each sub-frequency range in terms of their sample entropy, variance, skewness, kurtosis, and relative wavelet entropy. %The method is fully described in \DP{insert reference to Laurent's paper, or delete if useless}. 
% 

%The original detection algorithm, written in \textsc{Matlab}, was also translated into Python for flexible use and dissemination.}

\subsection{Time-delay embedding}

Pre-processing and segmentation yield a set of univariate timeseries, with an additional label corresponding to the type of event (background or seizure). Each time series then passed for analysis corresponds to the processed measurement of the electrical activity of a certain fish $f_i \in F$, belonging to one fish line $l_j \in L$, during a given event $e_k \in E$. Hence, each post-processed time series can be considered as an element $a_{i,j,k}$ of the tensor dataset $A = F \times L \times E$.

To analyze the dynamics of each $a_{i,j,k}$, we use a standard technique from dynamical systems theory: \textit{time-delay embedding}. The method has its theoretical grounds in Takens' theorem \cite{takens1981dynamical}, which provides the conditions to reconstruct a smooth attractor from observations made with generic functions. Its application is motivated by interpreting the fish's brain activity as a high-dimensional dynamical systems, of which the one-dimensional time series is a partial observation. The time-delay embedding allows us, to a certain extent, to reconstruct the full attractor associated with the fish' brain dynamics. 

In practice, given discrete time series (like the ones obtained from measurements) $\{x_0, x_1, \dots\}$, and evenly sampled time sequences $\{t_0, t_1, \dots\}$, the result of the embedding is a set of vectors in a $d$-dimensional state space, $X = \{x_{t_i}, x_{t_i + \tau}, \dots, x_{t_i + (d-1)\tau}\}$, with $i=0,1,\dots$. Each data point in $X$ represents the state of the reconstructed brain dynamics in $d$ dimensions at a given time. For example, processing the time series of a sinusoidal oscillation with time-delay embedding results in points forming a circle in a 2-dimensional space. Hence, the shape of the final embedding contains information about the type of dynamics considered and allows functional observability of dynamical systems \cite{montanari2022functional}. 

The method involves two parameters: the delay $\tau$, and the embedding dimension $d$. Both parameters are determined using standard techniques \cite{takens1981dynamical, kennel1992determining} implemented in the Python toolbox \texttt{giotto-tda} \cite{tauzin2020giottotda}. The same toolbox is then used to determine the time-delay embedding of each processed time series. 

\subsection{Topological data analysis}

\textit{Topological data analysis} as a field employs concepts from topology to analyze data. Topology provides mathematical tools to study and compare the shape properties of objects (or data), like connected components, and holes in different dimensions. It thus extract the topological bases of each shape, similarly to using basis vectors and corresponding rank to compare the structure of matrices \cite{smith2021topological}. A visual example is that of a donut, which is topologically equivalent to a mug because both display one hole. TDA allows to study topological properties in clouds of data points, sampled from dynamical attractors. In particular, the \textit{persistent homology} \cite{edelsbrunner2002topological, fugacci2016persistent} technique quantitatively describes the existence, or persistence, of these holes across several spatial resolution scales, and is robust against noise in the data.

Computing persistent homology consists of two main steps. First, it connects data points to form a \textit{simplicial complex}, a mathematical object which generalises the notion of graphs \cite{salnikov2018simplicial}. Simplicial complexes are defined as a set of nodes -- the data points -- and a set of simplices, which represent the connections between two or more nodes. For example, a simplex connecting nodes 1, 2, and 3 can be regarded as a $\{1, 2, 3\}$ object. 
This step is usually performed with the Vietoris-Rips process \cite{carlsson2006algebraic}: after embedding data points in a suitable space, and given a set of radii $\{r_1, r_2, \dots \}$ generating balls $\{ \mathcal{B}(r_1), \mathcal{B}(r_2), \dots \}$ around each data point, the procedure connects the nodes whose balls $\mathcal{B}(r_m)$ intersect. Each radius $r_m$ corresponds to a resolution scale for analysis: for a very small $r_m$, none of the points will be connected, whereas a large $r_m$ guarantees that all nodes are connected with all the others. Persistent homology is a multi-scale tool which creates a set of simplicial complexes, each one corresponding to a certain $r_m$.

Then, the persistent homology pipeline outputs a bardcode diagram, or equivalently a persistence diagram \cite{zomorodian2004computing}, which describes the data topology over the resolution scales. Both diagrams encode the radii at which a given-dimensional hole persists, marking the ``birth'' radius $r_B$ and ``death'' radius $r_D$ (respectively, the first and last radius that determine a simplicial endowed with a certain hole). The dimension of homology $H_n$ counts the number of $n$-dimensional holes: $H_0$ correspond to the number of connected components, $H_1$ count the number of empty circles (like in the donut case), $H_2$ counts the number of empty spheres, and so on. Here, we focus on $H_0$ and $H_1$. Holes with a longer ``lifetime'', that is,  more persistent across scales, are typically considered more important, whereas those with short lifetimes are considered noise. 

Finally, each persistence diagram obtained by the procedure above is summarised by extracting three topological metric: persistent entropy, total persistence, and persistent Betti numbers. For a persistence diagram $D=\{(r_{B_n}, r_{D_n}) \}_{n \in N}$, with $r_{D_n} < \infty$, the persistence entropy is defined as
\begin{equation}
    E(D) = - \sum_{n \in N} p_n \log (p_n) \, ,
    \label{eq:pers_entr}
\end{equation}
where $p_n = (r_{D_n} - r_{B_n})/ L_D$. Here, $L_D = \sum_{n \in N} (r_{D_n} - r_{B_n})$ is the total persistence. Intuitively, Eq. \ref{eq:pers_entr} is a measure of the entropy of each point in the diagram. The Betti number $\beta_n$ counts the number of unique n-holes in
a simplex. It can can be seen as the total number of unique topological features. While persistent Betti numbers and total persistence are arguably more well-known \cite{cerri2013betti}, we also employ persistent entropy since it is stable, scale invariant and more robust to noise \cite{atienza2020stability}. 

In the present analysis, we apply persistent homology on the data clouds generated by time-delay embedding, and use all three topological measures. For this task, we use functions from the \texttt{giotto-tda} Python package \cite{tauzin2020giottotda}.

\subsection{Statistical analysis}

Permutation analyses of variance (ANOVAs) performed and permutation $t$-tests were all performed with the standard value of permutations (5000 vs 9999, respectively), unless otherwise stated. Posthoc analysis was performed when applicable with pairwise Welch's $t$-tests, and Bonferroni-corrected when applicable.

\section*{Acknowledgments}
The authors would like to thank the Aquatic Platform of the LCSB for excellent animal handling. M.L. is a
Postdoctoral Researcher of the Fonds de la Recherche Scientifique–FNRS. C.D. and A.S.'s work was supported by the Luxembourg National Research Fund (FNR) through INTER/DFG/17/11583046 and the Caisse Médico-Complémentaire Mutualiste Luxembourg (CMCM) fellowship for C.D. in the framework of the Luxembourg National Research Fond (FNR) PRIDE DTU CriTiCS (grant reference 10907093). A.S. and D.F. acknowledge financial support of the Institute for Advanced Studies of the University of Luxembourg through an Audacity Grant (IDAE-2020). D.P. was partially funded by the European Union through the ERC INSPIRE grant (project number 101076926). Views and opinions expressed are however those of the author only and do not necessarily reflect those of the European Union or the European Research Council. Neither the European Union nor the granting authority can be held responsible for them. 
%dr. Cristina Donato for her valuable insight on seizure labelling. \DP{funding bodies}

\section*{Author contributions}
Conceived and designed the work: D.P, M.L., A.S.
Collected and curated data: C.D., D.F., A.S., D.P.
Developed software and performed computational analysis: M.L., D.F., L.M.
Interpreted results: D.P., M.L., D.F., A.S. Supervised the project: D.P., A.S. Wrote the paper: D.P., M.L., D.F., A.S.

\section*{Competing interests}
The authors declare no competing interests.

\section*{Data availability}
Original data are available upon request.

 \section*{Code availability}
 The code for seizure detection is publicly available at \url{https://github.com/Espritdelescalier/SeizureDetectionTool}. %\DP{github repository with code}.

\bibliography{biblio}

\clearpage

\newcommand\SupplementaryMaterials{%
  \xdef\presupfigures{\arabic{figure}}% save the current figure number
  \xdef\presupsections{\arabic{section}}% save the current section number
  \renewcommand{\figurename}{Supplementary Figure}
  \renewcommand\thefigure{S\fpeval{\arabic{figure}-\presupfigures}}
  \renewcommand\thesection{S\fpeval{\arabic{section}-\presupsections}}
}

\SupplementaryMaterials

\onecolumngrid
\setcounter{page}{1}
\renewcommand{\thepage}{S\arabic{page}}
\setcounter{equation}{0}
\renewcommand{\theequation}{S\arabic{equation}}

\begin{center}
{\Large\bf Supplementary Information}\\[5mm]
%{\large{Hypergraphs or simplicial complexes: Untangling the effect of higher-order representations on collective dynamics}}\\[1pt]
\end{center}

\begin{figure}[htb]
    \centering
    \includegraphics[width=.55\linewidth]{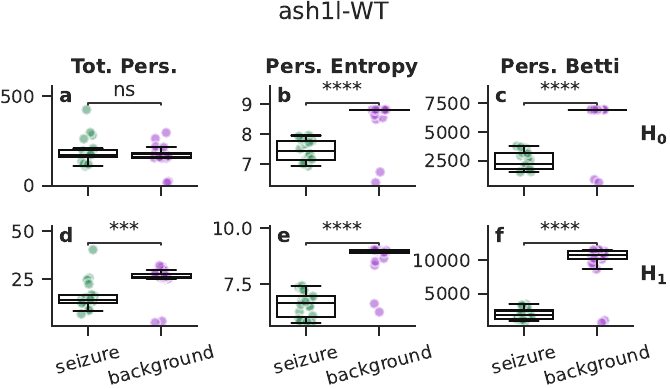}

    \includegraphics[width=.55\linewidth]{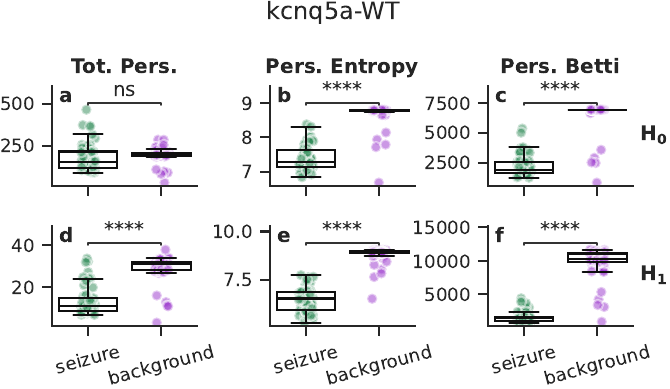}

    \includegraphics[width=.55\linewidth]{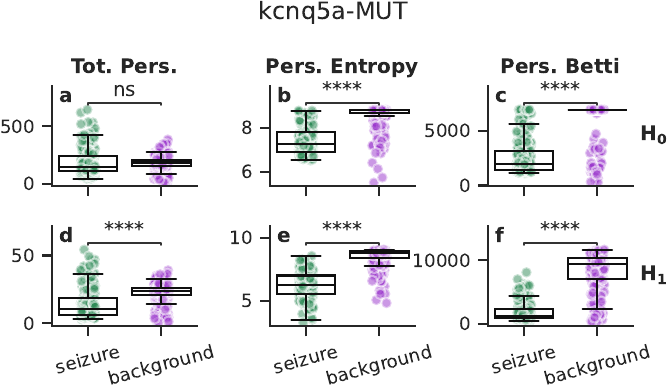}
    \caption{\textbf{The topology of dynamics discriminates between seizures and background LFP activity.} In complement to \cref{fig:2}, here we show three more subgroups: (ash1l, WT), (kcnq5a-WTà, and (kcnq5a-WT). In each case, we show three topological metrics---(a, d) total persistence, (b, e) persistent entropy, and (c, f) persistent Betti---for seizures and background, in homological dimensions (a-c) 0 and (d-f) 1. Pairwise Welch's $t$-test indicate a significant difference in all cases with $p<10^{-15}$ (``****''), except for total persistence in $H_0$ where the difference is not significant (``ns''). }
    \label{fig:s1}
\end{figure}

\begin{figure}[htb]
    \centering
    \includegraphics[width=.6\linewidth]{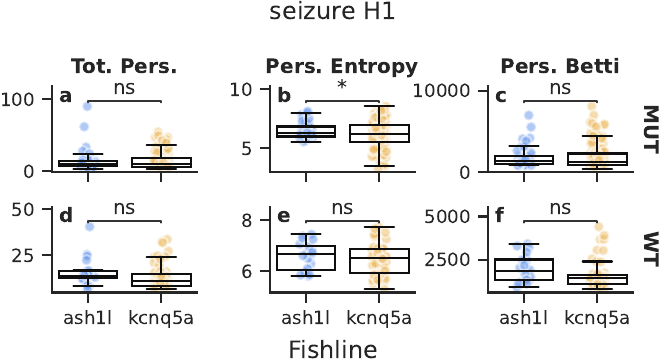}

    \caption{\textbf{Different fishlines display similar topological patterns at $H_0$ but can be discriminated from background at $H_1$}. We show three topological metrics---total persistence, persistent entropy, and persistent Betti---for two fishlines---ash1l and kcnq5a---in homological dimensions 0 and 1. Pairwise Welch's $t$-test indicate no significant (``ns'') difference in $H_0$ but persistent entropy and persistent Betti significantly differ between fishlines (``**'').}
    \label{fig:s2}
\end{figure}

\begin{figure}
    \centering
    \includegraphics[width=0.6\linewidth]{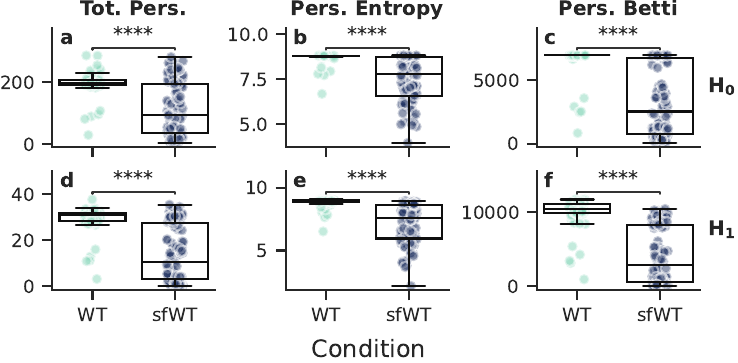}
    \caption{Same figure as \cref{fig:4}, but showing the results of pairwise Levene tests, indicating that the sfWT group has significantly larger variance, with $p<10^{-15}$ (``****'')}
    \label{fig:s3}
\end{figure}

\begin{figure}
    \centering
    \includegraphics[width=0.6\linewidth]{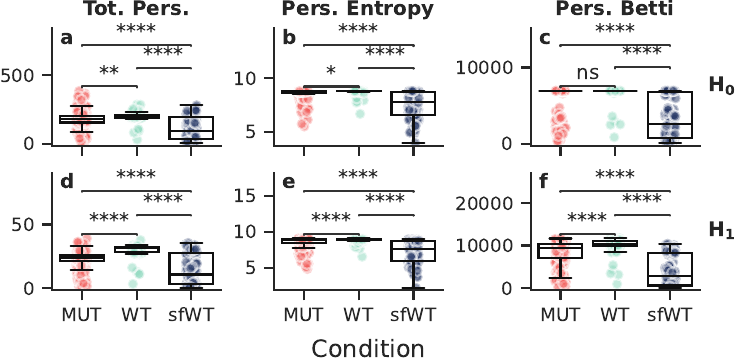}
    \caption{Same figure as \cref{fig:4}, but here we add the MUT group.}
    \label{fig:s4}
\end{figure}

% \begin{table}[ht]
%     \centering
%     \caption{Significant Effects from Three-way ANOVA for $H_0$ and $H_1$}
%     \begin{tabular}{lllr}
%         \toprule
%         \textbf{Metric} & \textbf{Factors} & \textbf{$p$-value} & \textbf{ES (\%)} \\
%         \midrule
%         \multicolumn{4}{l}{{$H_0$}} \\
%         \midrule
%         Total Persistence & & & \\
%         & Triple Interaction & 0.02 & 0.4 \\
%         \cmidrule(r){2-4}
%         Persistent Entropy & & & \\
%         & Event & $<10^{-15}$ & 58.0 \\
%         & Triple Interaction & 0.010 & 0.7 \\
%         \cmidrule(r){2-4}
%         Persistent Betti & & & \\
%         & Event & $<10^{-15}$ & 58.0 \\
%         & Triple Interaction & $<10^{-15}$ & 0.8 \\
%         \midrule
%         \multicolumn{4}{l}{$H_1$} \\
%         \midrule
%         Total Persistence & & & \\
%         & Event & $<10^{-15}$ & 24.2 \\
%         & Condition & $<10^{-15}$ & 1.4 \\
%         & Condition:Fishline & 0.031 & 0.5 \\
%         & Triple Interaction & $<10^{-15}$ & 1.3 \\
%         \cmidrule(r){2-4}
%         Persistent Entropy & & & \\
%         & Event & $<10^{-15}$ & 66.6 \\
%         & Fishline & 0.045 & 1.3 \\
%         & Condition:Fishline & 0.030 & 0.6 \\
%         \cmidrule(r){2-4}
%         Persistent Betti & & & \\
%         & Event & $<10^{-15}$ & 74.4 \\
%         & Fishline & $<10^{-15}$ & 2.0 \\
%         & Condition:Fishline & 0.033 & 0.7 \\
%         & Triple Interaction & $<10^{-15}$ & 1.2 \\
%         \bottomrule
%     \end{tabular}
% \end{table}

\end{document}